
\overfullrule=0pt
\magnification=\magstep1
\baselineskip=5ex
\raggedbottom
\font\fivepoint=cmr5
\headline={\hfill{\fivepoint VBEHLMLJPS -- 16/July/93}}
\def\r{{\bf r}}
\def\HF{{\rm HF}}
\def\Det{{\rm Det}}

\def\uprho{\raise1pt\hbox{$\rho$}}
\def\mfr#1/#2{\hbox{${{#1} \over {#2}}$}}
\font\eightpoint=cmr8
\def\upvarphi{\raise1pt\hbox{$\varphi$}}
\centerline{\bf THERE ARE NO UNFILLED SHELLS IN HARTREE-FOCK THEORY}

\vskip 1truecm
{\baselineskip=3ex
\centerline{Volker Bach$^{(1),}$\footnote{$^{*}$}{\baselineskip=12pt
\eightpoint  Current
address: FB Mathematik MA 7-2, Technische Universit\"at Berlin, Strasse des
17 Juni 136, D-W-1000 Berlin, Germany},
Elliott H. Lieb$^{(1),(2),}$, Michael Loss$^{(3)}$
and
Jan Philip Solovej$^{(2),}$
\vfootnote{}{\eightpoint \copyright 1993 by the authors.
Reproduction of this article by any means is permitted for non-commercial
purposes.}}}
\bigskip
\noindent{\it $^{(1)}$Department of Physics, Jadwin Hall, Princeton
University, P.O. Box 708, Princeton, NJ, 08544\hfil\break
$^{(2)}$Department of Mathematics, Fine Hall, Princeton University,
Princeton, NJ, 08544\hfil\break
$^{(3)}$School of Mathematics, Georgia Institute of Technology, Atlanta, GA
30332}
\bigskip
\bigskip
{\narrower{\it Abstract:\/} Hartree-Fock theory is supposed to yield a
picture of atomic shells which may or may not be filled according to the
atom's position in the periodic table.  We prove that shells are always
completely filled in an exact Hartree-Fock calculation.  Our theorem
generalizes to any system having a two-body interaction that, like the
Coulomb potential, is repulsive.\smallskip}
\bigskip
\bigskip

The Hartree-Fock (HF) variational calculation provides an approximate
determination of ground states and ground state energies of quantum
mechanical systems such as atoms and molecules, and is widely used in
physics and chemistry.

The picture of atoms it is supposed to yield is one of ``shells'' (or
degenerate eigenvalues of the HF operator) which may or may not be filled
according to the position of the atom in the periodic table.  (Indeed, the
concept of shells stems from HF theory itself, for shells are not
overwhelmingly evident, or even precisely defined in the exact many-body wave
function.)  Despite years
of attention to this subject, it is surprising that this notion of unfilled
shells has gone unchallenged.  A very simple proof, which we give here,
shows that {\it shells are always filled in HF theory}.  In other words,
the degeneracy of the
highest filled level of the HF operator is always exactly what is needed to
accommodate the assumed number of electrons.

The theorem and its proof given below obviously generalize to any
system in which the two-body interaction $V$ is repulsive, i.e.,
positive definite as an operator on the two-particle Hilbert space.  In
particular, $V$ is allowed to be spin dependent and to contain
projection operators, as in the nuclear physics setting.  The
electronic Coulomb repulsion, for example, satisfies this positivity
condition.  The one-body part of the Hamiltonian can be arbitrary.  For
convenience and because of its familiarity, we use the atomic
Hamiltonian as an illustration.

Thus, we consider a Hamiltonian
$$H = \sum \limits^N_{i=1} \left( - {\hbar^2 \over 2m} \Delta_i + U
(\r_i)\right) + \mfr1/2 \sum \limits_{i \not= j} V(\r_i, \r_j)$$
acting on $N$-electron wave functions, i.e. wave functions $\Psi (\r_1,
\sigma_1; \dots ; \r_N, \sigma_N)$ that are antisymmetric with respect to
interchanging $(\r_i, \sigma_i)$ with $(\r_j, \sigma_j)$.  In the example
of an atom with nuclear charge $Ze, \,U$ and $V$ would be given by $U(\r) = -
Ze^2/r$ and $V (\r, \r^\prime) = e^2 \vert \r - \r^\prime \vert^{-1}$.

To obtain an approximate value for the ground state energy, $E_Q = \min
\langle \Psi \vert H \vert \Psi \rangle /\langle \Psi \vert \Psi \rangle$,
the HF calculation restricts attention to Slater determinants, i.e., wave
functions of the form
$$\Phi (\r_1, \sigma_1; \dots ; \r_N, \sigma_N) = (N!)^{-1/2} \Det \{ f_i
(\r_j, \sigma_j) \} \eqno(1)$$
in which $f_1, \dots , f_N$ are orthonormal functions of space and spin:
$\langle f_i \vert f_j
\rangle = \delta_{ij}$.  The approximate ground state energy is then given
by the HF-energy,
which is defined to be
$$E_{\HF} = \min \langle \Phi \vert H \vert \Phi \rangle, \quad \Phi
\ \hbox{has the form (1)}. \eqno(2)$$
Any minimizer, $\Phi_{\HF}$, i.e., a determinantal function satisfying
$E_{\HF} = \langle
\Phi_{\HF} \vert H \vert \Phi_{\HF} \rangle$, is a HF ground state.
It may not be unique.
We remark that mathematical precision actually requires an ``infimum''
rather than ``minimum'' in (2) because a HF ground state may not exist.
This will be the case, e.g., for an atom with $N > 2Z + 1$, i.e., a very
negative ion [1].  For neutral or positively ionized atoms and
molecules, however, it was proved in [2] that a HF ground state does exist
and, at least in this case, the word ``minimum'' in (2) is justified.

If a HF ground state does exist, it necessarily obeys the HF (or
self-consistent field) eigenfunction equations
$$h_\Phi \upvarphi_k = \varepsilon_k \upvarphi_k \eqno(3)$$
for all $1 \leq k \leq N$, where $h_\Phi$ is the one-body operator
defined by its action on an arbitrary function of one space-spin
variable by
$$\eqalign{(h_\Phi f) (\r, \sigma) = &\left( - {\hbar^2 \over 2m}
\Delta + U(\r) + \int \sum \limits_{\tau = \pm 1} \sum \limits^N_{j=1}
\vert \upvarphi_j (\r^\prime, \tau) \vert^2 V(\r, \r^\prime) d^3
\r^\prime \right) f(\r, \sigma) \cr &- \sum\limits_{\tau = \pm 1} \sum
\limits^N_{j=1} \upvarphi_j (\r, \sigma) \int \overline{\upvarphi}_j
(\r^\prime, \tau) f(\r^\prime, \tau) V(\r, \r^\prime) d^3 \r^\prime,
\cr} \eqno(4)$$
and where $\upvarphi_1, \dots , \upvarphi_N$ denote the special $N$
orthonormal functions comprising the energy minimizing Slater
determinant $\Phi_{\HF}$.  The eigenvalues, $\varepsilon_k$, of
$h_\Phi$ give us some insight into the possible energy levels for
binding an extra electron, but that is not our concern here.

{\bf Theorem:}  {\it Assume that $V$ is positive definite, i.e., for every
nonzero function $\psi$ of two space-spin variables
$$\sum \limits_{\sigma_1, \sigma_2 = \pm 1} \int \vert \psi (\r, \sigma ;
\r^\prime, \sigma^\prime) \vert^2 V(\r, \r^\prime) d^3 r d^3 r^\prime > 0.$$
Let $\upvarphi$ be an eigenfunction
of $h_\Phi$ with eigenvalue $\varepsilon$ (i.e., $h_\Phi
\upvarphi = \varepsilon \upvarphi)$ that is orthogonal to
the minimizing set $\upvarphi_1, \dots , \upvarphi_N$, i.e.,
$\langle \upvarphi
\vert \upvarphi_k \rangle = 0$ for all $1 \leq k \leq N$.  Then
$\varepsilon > \varepsilon_k$ for all $1 \leq k \leq N$.}

Before proving this theorem let us point out its main corollaries.
First, it implies that the functions $\upvarphi_1, \dots , \upvarphi_N$
comprising $\Phi_{\HF}$ occupy the $N$ {\it lowest\/} energy levels of
$h_\Phi$; the reader may or may not find this surprising, but we point
out that there is no proof of this assertion without the assumption
that $V \geq 0$.  {\it Our main point, however, is the second
implication which, indeed, is surprising:  $\Phi_{\HF}$ does not leave
any degenerate level unfilled.  There is a gap because $\varepsilon >
\varepsilon_k$ for all $k = 1,2, \dots , N$.}

{\it Proof of the theorem:}  Assume $\varepsilon_1 \leq \varepsilon_2 \leq
\dots \leq \varepsilon_N$.  We shall derive a contradiction to the
assumption that $\varepsilon \leq \varepsilon_N$.  First, we
introduce some more notation.  Denote $\varepsilon$ by $\varepsilon_{N+1}$
and $\upvarphi$ by $\upvarphi_{N+1}$.  Further for all $1 \leq k
\leq N + 1$ and $1 \leq l \leq N + 1$, we
define
$$\eqalign{h_k &= \left\langle \upvarphi_k \biggl\vert - {\hbar^2 \over 2m}
\Delta - U(\r)\biggr\vert \upvarphi_k \right\rangle \quad \hbox{and} \cr
V_{k,l} &= \sum \limits_{\sigma, \sigma^\prime = \pm 1} \int \mfr1/2
\big\vert \upvarphi_k (\r, \sigma) \upvarphi_l (\r^\prime, \sigma^\prime) -
\upvarphi_l (\r, \sigma) \upvarphi_k (\r^\prime, \sigma^\prime) \big\vert^2
V(\r, \r^\prime) d^3 r d^3 r^\prime .\cr}$$
Notice that $V_{k,k} = 0$ and $V_{k,l} > 0$ if $k \not= l$ since $V$ is
positive definite.

Now let $\widetilde \Phi$ be the Slater determinant built from $\upvarphi_1,
\dots \upvarphi_{N-1}, \upvarphi_{N+1}$, as in (1).  One easily checks that
$$\eqalign{\langle \Phi_{\HF} \vert H \vert \Phi_{\HF} \rangle &= \sum
\limits^N_{k=1} h_k + \mfr1/2 \sum \limits^N_{k,l=1} V_{k,l}\ \ , \cr
\langle \widetilde \Phi \vert H \vert \widetilde \Phi \rangle &= \sum
\limits^{N-1}_{k=1} h_k + \mfr1/2 \sum \limits^{N-1}_{k,l=1} V_{k,l} + h_{N+1}
+ \sum \limits^{N-1}_{l=1} V_{l,N+1}\ \ , \cr}$$
and, for $1 \leq k \leq N + 1$,
$$h_k + \sum \limits^N_{l=1} V_{k,l} = \langle \upvarphi_k \vert h_{\HF} \vert
\upvarphi_k \rangle = \varepsilon_k .\eqno(5)$$
Notice that the term $l = k$ in the sum in
(5) does not contribute since $V_{k,k} = 0.$

Since $\Phi_{\HF}$ is the HF ground state it follows that $\langle \Phi_{\HF}
\vert H \vert \Phi_{\HF} \rangle \leq \langle \widetilde \Phi \vert H \vert
\widetilde \Phi \rangle$ and, thus,
$$\eqalign{0 &\leq \langle \widetilde \Phi \vert H \vert \widetilde
\Phi \rangle - \langle \Phi_{\HF} \vert H \vert \Phi_{\HF} \rangle \cr
&= h_{N+1} - h_N + \sum \limits^{N-1}_{l=1} (V_{l,N+1} - V_{l,N}) \cr
&= \varepsilon_{N+1} - \varepsilon_N - V_{N,N+1} \leq - V_{N,N+1}. \cr}$$
The last inequality uses the assumption $\varepsilon_{N+1} \leq \varepsilon_N$,
but we then have the contradiction $0 \leq - V_{N,N+1} < 0$.  QED

The proof does not give a rigorous estimate of the gap $\varepsilon_{N+1} -
\varepsilon_N$, but it does show that the gap is at least $V_{N,N+1}$,
which is usually not a tiny quantity.  Thus, even an ``approximate''
degeneracy is unlikely.

This work was supported by the U.S. National Science Foundation through the
following grants:  PHY90-19433 A02 [VB and EHL]; DMS92-07703 [ML];
DMS92-03829 [JPS].
\bigskip
\bigskip\noindent
{\bf References}
\item{[1]}  E.H. Lieb, {\it Bound on the maximum negative ionization of
atoms and molecules}, Phys. Rev. {\bf 29A}, 3018 (1984).
\item{[2]}  E.H. Lieb and B. Simon, {\it The Hartree-Fock theory for
Coulomb systems}, Commun. Math. Phys. {\bf 53}, 185 (1977).  See also E.H.
Lieb, {\it Thomas-Fermi and Hartree-Fock theory} in Proceedings of the 1974
International Congress of Mathematicians, vol. 2, p. 383 (1975).

\bye